\documentclass[11pt]{article}
\usepackage[a4paper, portrait, margin=2.5cm]{geometry}
\usepackage{graphicx}
\usepackage{lastpage}
\usepackage{fancyhdr}
\usepackage{layout}

\usepackage[backend=biber,style=numeric]{biblatex}
\fancypagestyle{firstpage}{%
  \setlength\headheight{154pt}
  \fancyhf{}%
  \fancyhead[C]{\includegraphics[width=\textwidth]{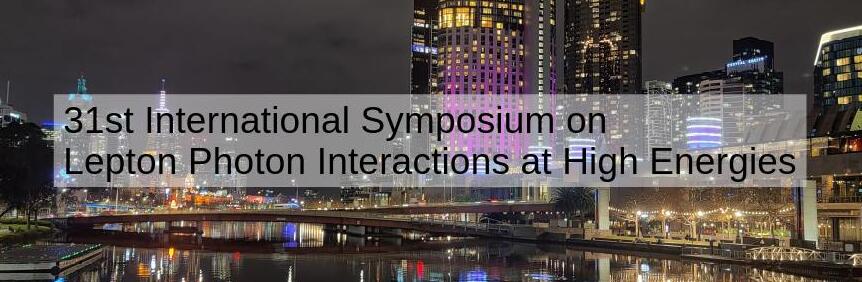}}%
  \fancyfoot[C]{CC-BY-4.0 licence}%
}

\begin{document}
\title{Systems for detecting and measuring backgrounds with the SABRE South experiment} 
\date{17-21 July 2023}
\author{L. J. Milligan \\ School of Physics, University of Melbourne \\ ARC Centre of Excellence for Dark Matter Particle Physics \\ \textit{On behalf of the SABRE South Collaboration}} 

\newgeometry{top=2cm, bottom=7cm}
\maketitle
\thispagestyle{firstpage}
\begin{abstract}
The SABRE (Sodium iodide with Active Background REjection) experiment aims to detect an annual rate modulation from dark matter interactions in ultra-high purity NaI(Tl) crystals which will provide a model independent test of the signal observed by DAMA/LIBRA. SABRE will consist of two separate detectors in the Northern and Southern hemispheres. SABRE South will be located in the newly completed Stawell Underground Physics Laboratory (SUPL), the first deep underground laboratory in the southern hemisphere. The combination of SABRE North and South is intended to disentangle seasonal or site-related effects from the dark matter-like modulated signal. Measuring and understanding backgrounds is essential for the reliability and consistent performance of these searches, and as the first large detector in SUPL SABRE South will also be used to measure backgrounds from radiogenic and cosmogenic sources. The SABRE South veto system is designed to detect the signals generated by radiation and cosmic rays using a 12 kL linear alkyl-benzene (LAB) based liquid scintillator (LS) detector contained in a steel vessel and instrumented with 18 Hamamatsu R5912 photomultiplier tubes (PMTs), alongside a plane of 8 plastic scintillator modules (instrumented with 2 R13089 PMTs) located above the vessel to reliably detect muons from cosmic-rays with a position resolution of 5 cm.

\end{abstract}

\newpage
\restoregeometry

\section{Introduction}

Dark matter (DM) is a hypothesised form of matter proposed to explain a wide range of otherwise unexplainable astrophysical observations. Numerous experiments have been established in the past decades that aim to detect particle dark matter via some form of nuclear scattering. An early result from this field of direct detection comes from the DAMA/LIBRA experiment, which searched for a modulating signal in their data --- an annually modulating flux with a period of 1 year peaking in June, due to the relative motion of the Earth within the galactic DM halo. A signal with these properties has been observed by DAMA for 25 years at a significance of 12.9 $\sigma$ \cite{Bernabei:2020mon}, but is in strong tension with other direct detection experiments making use of different target materials. A conclusive test of the DAMA result therefore requires a model independent reproduction of the measurement using the same target material. The SABRE experiment aims to perform such a test, leveraging, among other things, active background rejection, and dual hemisphere data. Whilst dual hemisphere data allows for disentanglement of seasonal effects via the comparison of data from the SABRE North (Laboratori Nazionali del Gran Sasso, LNGS) and the SABRE South (Stawell Underground Physics Laboratory, SUPL) detectors, a detailed understanding of any background contributions in the data is desirable and can be obtained from our veto systems. These systems in combination will allow us to identify and veto the contributions to any potential seasonal or time varying background effect --- such as muon induced neutron spallation, or time-dependent detector effects --- thus complementing the dual hemisphere comparison. 


\section{The SABRE South Veto Systems}

\begin{figure}
    \centering
    \includegraphics[width=0.6\columnwidth]{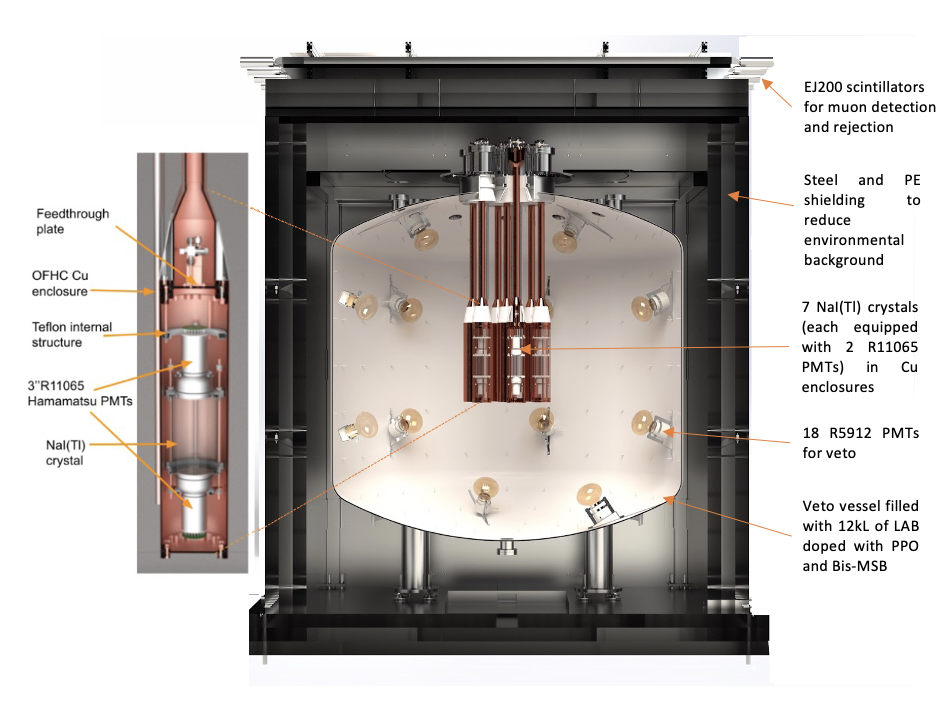}
    \caption{A render of the complete SABRE South detector. The active LS veto surrounds the NaI(Tl) detectors, with the muon veto placed on top of the shielding.}
    \label{fig:detector}
\end{figure}

Labelled in Figure \ref{fig:detector}, the liquid scintillator (LS) veto detector consists of 12 kL of a LAB based liquid scintillator, which is contained in a steel vessel that is covered internally in lumirror, and instrumented with 18 Hamamatsu 20cm R5912 photomultiplier tubes (PMTs). The muon veto system is placed above the shielding as shown in Figure \ref{fig:detector}, and consists of a plane of 8 plastic scintillator modules (each instrumented with 2 Hamamatsu R13089 PMTs) placed side-by-side, where each muon detector is 3~m$\times$0.4~m, together providing a 9.6~m$^2$ coverage above the detector. This proceeding concentrates on the veto detectors, but more detail regarding the crystal detectors and background modelling can be in found in \cite{SABRE:2022twu, tdr}. To effectively complement the crystal detectors, the muon and LS detectors will need to be able to effectively detect and veto backgrounds for the lifetime of the experiment, and demonstrate a capability for background reconstruction. 

\subsection{The LS Active Veto}

The LS veto needs to continuously meet three main requirements through the lifetime of the detector: (i) the tagging and rejection of backgrounds which deposit greater than 50 keV$_{ee}$ within the LS, (ii) a factor of 10 reduction for the peaking $^{40}$K background from 0.1 to 0.01 cpd/kg/keV$_{ee}$ (peaking within the signal ROI of 1-6 keV$_{ee}$) \cite{SABRE:2022twu} via the tagging of a coincident gamma ray observed with the LS veto, and (iii) the ability to reconstruct event/background mechanisms which may mimic a DAMA-like annual modulation signal --- i.e. muon-induced neutron spallation, and other time dependent backgrounds. 

\begin{figure}[!h]
    \centering
    \includegraphics[height=5.5cm]{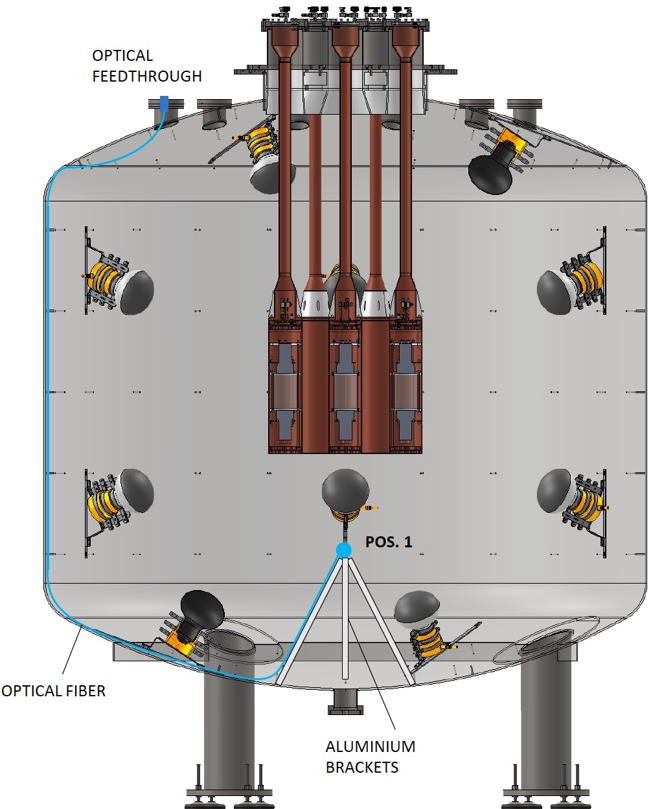}
    \includegraphics[height=6.5cm]{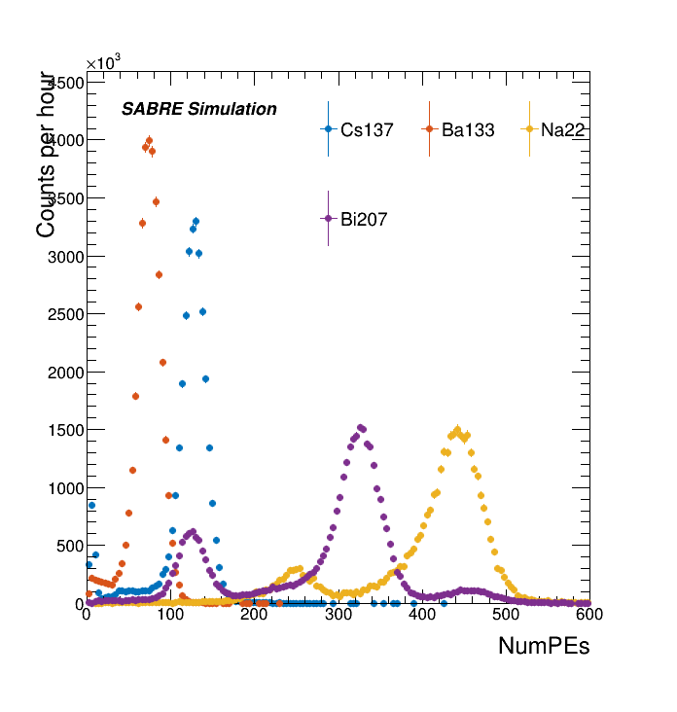}
    \caption{(Left) The proposed design for the in-situ optical calibration system. (Right) The number of PEs detected by the LS veto detector for the energy depositions corresponding to each source.}
    \label{fig:VetoCal}
\end{figure}

To continuously fulfill said requirements, accurate monitoring of the detector and PMT performance over time is required, particularly the PMT gain stability and energy scale reconstruction. Lack of control/monitoring over said detector parameters can result in increased detector noise passing threshold requirements or decreased veto efficiencies, as thresholds are not adjusted to account for drifts/fluctuations in PMT gain or detector performance --- directly impacting the ability of the veto system to meet its background rejection requirements. In-situ calibration systems have been designed to ensure the accurate monitoring of performance over time. Two calibration systems will be used: (i) a light injection calibration system designed to specifically monitor PMT performance/stability, and (ii) a radioactive calibration system designed to monitor overall detector performance and energy reconstruction. For the former, a 445 nm laser will be injected into the system via an optical fibre attached to a diffusion light-bulb mounted on a tripod at the bottom of the vessel. Light will be periodically injected into the system in order to perform in-situ gain calibration of the PMTs. Figure \ref{fig:VetoCal} (Left) demonstrates how the system may be installed in practice. The radioactive calibration system operates via the lowering of radioactive sources into calibration tubes attached to the outer flanges of the vessel, placed in alternating positions to allow proper coverage of the vessel interior. The sources (gamma ray energy) to be used are $^{137}$Cs (662 keV), $^{133}$Ba (81 and 356 keV), $^{22}$Na (2 $\times$ 1275 keV), and $^{207}$Bi (570 and 1633 keV). Figure \ref{fig:VetoCal} (Right) plots the total number of photo-electrons (PEs) summed over all PMTs for the energy depositions corresponding to each source for a simulation of the calibration system.

\begin{figure}[!h]
    \centering
    \includegraphics[height=4.5cm]{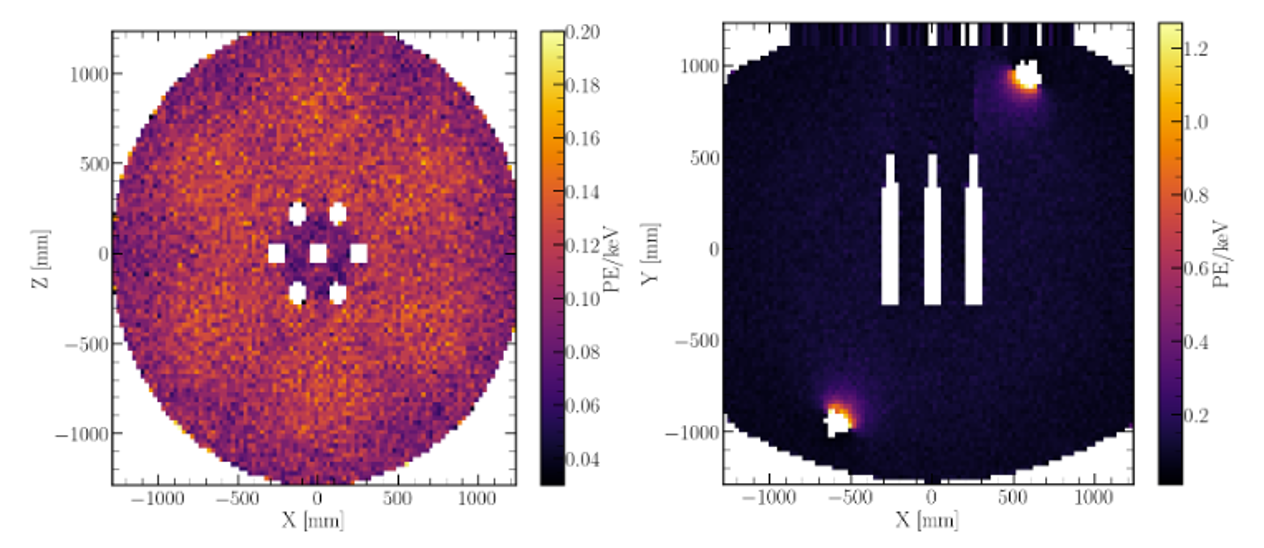}
    \caption{Probability heatmaps for PE detection probability within the LS veto. The right plot represents a top-down slice in the x-y plane, and the left shows a side-on slice in the x-z plane showing 2 of the 18 veto PMTs.}
    \label{fig:VetoPerformance}
\end{figure}

In addition to the development of in-situ calibration systems, studies have been undertaken to understand the detector's potential for event reconstruction. A probability heatmap can be constructed in order to understand PE detection probability, with optical physics simulations informing the geometric dependence of this probability. Details on the statistical calculations involved can be found in the SABRE South TDR \cite{tdr} and indicate an average PE detection probability of about $0.2$ PE/keV/PMT --- shown by the position dependent heat maps shown in Figure \ref{fig:VetoPerformance}.


\begin{figure}[!h]
    \centering
    \includegraphics[height=6cm]{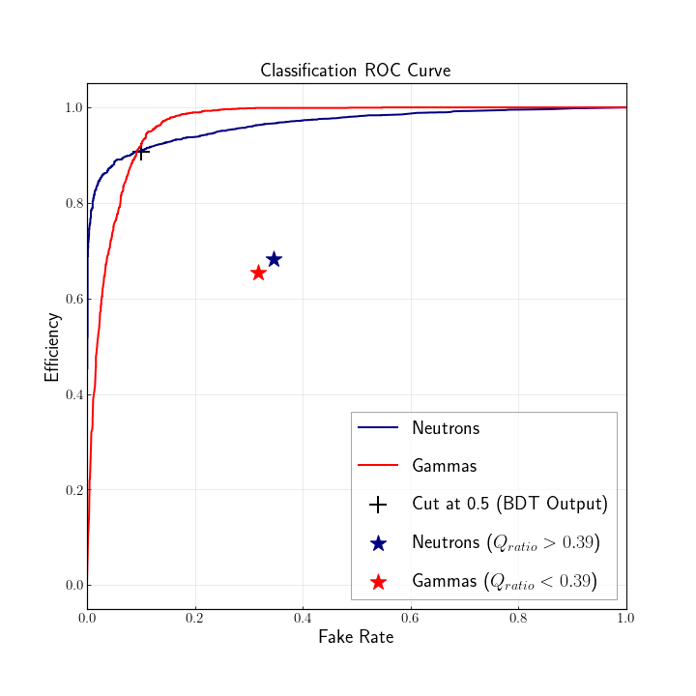}
    \includegraphics[height=6cm]{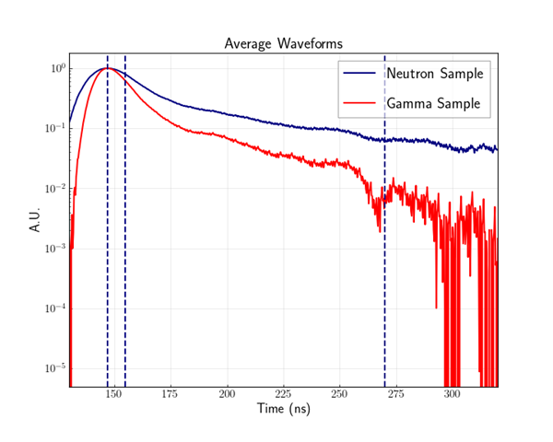}
    \caption{(Left) The ROC curve resulting from the BDT classification of neutron/gamma events in the small LS test vessel. (Right) Example average gamma/neutron pulses in the test vessel, with the prompt and delayed charge windows.}
    \label{fig:PID}
\end{figure}

Studies assessing the LS veto's potential for gamma ray and neutron discrimination were also performed. Utilising a small LS test vessel, containing approx. 40 L of LS and one PMT, pulse shape discrimination (PSD) variables are calculated for gamma rays from a $^{60}$Co source and neutrons originating from an Americium-Beryllium (Am-Be) source. Neutrons were selected via a time-of-flight cut between the test detector and one of the muon veto panels. Gamma rays from the $^{60}$Co source were selected with a charge cut. Four key variables are calculated: skew, kurtosis, a delayed/prompt charge ratio, and the amplitude-weighted mean time of the detected pulse. The charge ratio is defined as $Q_{ratio} = Q_{delayed}/Q_{prompt}$, with the windows used for $Q_{delayed}$ and $Q_{prompt}$ shown in Figure \ref{fig:PID} (Right). These four PSD variables were then fed into a boosted decision tree (BDT), using the XGBoost software package, in order to study their multivariate discrimination power. Figure \ref{fig:PID} (Left) shows the Receiver Operator Characteristic (ROC) curve of the BDT trained on pulses originating from $^{60}$Co gamma rays and Am-Be neutrons, and demonstrates great improvement over the use of a single standard PSD variable (in this case, the charge ratio variable). Since the small LS test detector is likely to have a higher PE detection efficiency compared to the LS veto, these results do not directly carry over. Additional simulation-based studies are being performed to extend this PID methodology to the final LS veto detector. A more detailed explanation of this prototype study will be outlined in a future SABRE South publication. 

\subsection{The Muon Veto}

\begin{figure}[!h]
    \centering
    \includegraphics[height=4.5cm]{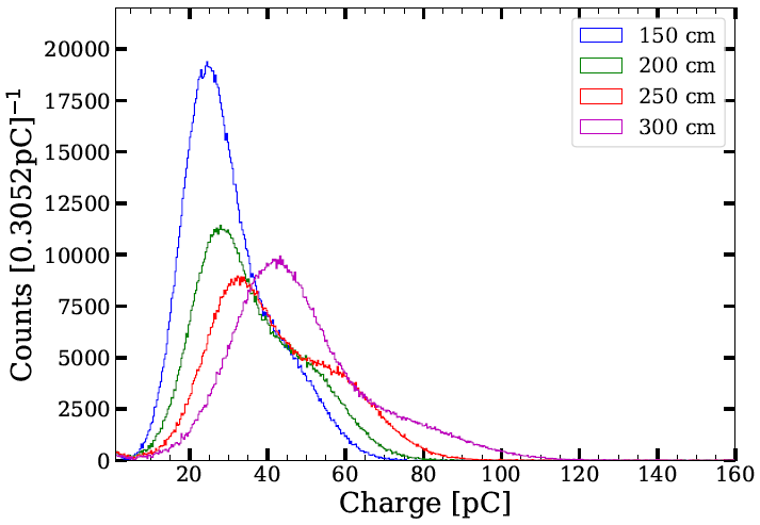}
    \includegraphics[height=5cm]{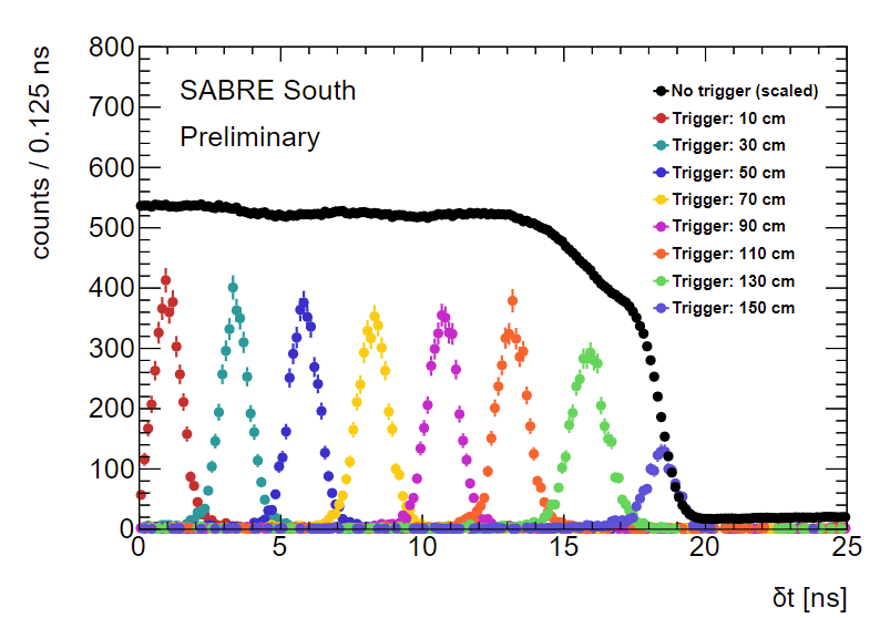}
    \caption{(Left) The charge histograms resulting from a $^{60}$Co calibration with the linear stage --- 150 cm is the centre of the panel. (Right) A plot of the time difference between the signals detected by each PMT on the muon panel, for different longitudinal positions of a $^{60}$Co placed on the panel --- 150 cm is defined as one end of the panel, with 0 cm the centre.}
    \label{fig:muons}
\end{figure}

Mirroring the LS veto requirements, the muon veto will need to accurately veto cosmogenic muons in the long-term, and maintain the ability to reconstruct backgrounds in partnership with the LS veto. To monitor the energy calibration, and by proxy the stability of the PMT gain, a radioactive calibration system has been developed utilising a linear stage capable of monitoring the response of each of the 8 muon detector panels, as well as the longitudinal response of individual panels. This ensures that thresholds can be adjusted in line with any drift in the energy response of the muon detectors, and that any change in the timing response of the detectors that may impact position reconstruction is accounted for. An example of the variation in the $^{60}$Co charge response along the panel is shown in Figure \ref{fig:muons} (Left), which can be used to check the stability of the timing response of detector, and thus it's stability with regards to position reconstruction.

Studies assessing the longitudinal position reconstruction of individual muon detectors have also been performed, utilising the relative difference in arrival time between signals detected by the PMTs at either end of the scintillator panels. Based on experiments where a $^{60}$Co was placed at regular intervals along the detector, it's found that the position of the source can be reconstructed with a position resolution of approximately 5 cm, with a timing resolution of 400 ps, shown in Figure \ref{fig:muons} (Right). This allows for accurate position reconstruction along one direction of the full muon veto system, with the other direction limited by the number of panels and their widths.

\section{Conclusion}

The SABRE South experiment will employ the capabilities of two detector systems for background detection/characterisation. Both systems will leverage in-situ calibration systems to ensure long-term monitoring of detector stability and operation, and accurate background reconstruction. Radioactive calibration systems are utilised in both vetos, whereas the LS veto also incorporates in-situ optical calibration. Background reconstruction techniques have been explored for both detectors, with a small-scale test detector demonstrating there is potential for PID in the full LS veto, and accurate longitudinal position reconstruction for the muon veto.

\printbibliography

\end{document}